\documentclass{article}
\usepackage{graphicx} 
\usepackage{amsmath}
\usepackage[affil-it]{authblk}
\usepackage{mathtools}
\usepackage{geometry}
\usepackage{float}
\newgeometry{top=3cm,left=3cm,right=3cm,bottom=3cm}
\usepackage{quantikz}
\usepackage{hyperref}
\usepackage{amssymb}
\usepackage{mhchem}
\usepackage{array}
\usepackage[super,sort,compress]{cite}
\usepackage{amsmath}
\usepackage{xcolor}
\usepackage{float}
\let\origfigure\figure
\let\endorigfigure\endfigure
\renewenvironment{figure}[1][H] {\origfigure[H]} {\endorigfigure}

\title{Teleportation with non-maximally entangled states and underlying unitary algebras of certain  bipartite systems  }

\author{P. Dasgupta \and D. Gangopadhyay }

\affil{ Department of Physics, Sister Nivedita University, DG Block , Newtown, Action Area 1, 
	
	Kolkata 700156, India }

\begin{document}

\maketitle

\begin{abstract}
New convenient thumbrules are obtained to test entanglement of wavefunctions for bipartite qubit and qutrit systems. All  results are analytic. The new results are :(a)For bipartite qubit systems there exists a matrix $A$ for which  $Det A =0$ implies unentanglement while $Det A\neq 0 $ implies entanglement. There is an underlying SU(2) algebra. (2)Teleportation for a general qubit state is possible by using non-maximally entangled bipartite qubit states. This protocol has an additional parameter , {\it viz.}, $Det A$, which enhances the cryptographic security of the teleportation. (c) For qutrits there is a matrix $P$ for which $Det P =0 $ simultaneously with $Tr P = \pm 1$ imply unentanglement. Any departure  from these conditions implies entanglement. There exists an underlying SU(3) algebra. (d) Physical interpretation of the underlying algebras are given and plausible experimental scenarios are proposed for the SU(2) case in the context of two entangled electrons. 
(e) The entanglement entropy in both cases , {\it viz.} for qubits and qutrits respectively, are expressed in terms of  the determinants and trace of the matrices mentioned above.
\end{abstract}

\markboth{P.Dasgupta and D.Gangopadhyay}
{Teleportation with non-maximally entangled states }

\section{Introduction} 
	The concept of entanglement  materialized first in the papers$^{1-3}$. Quantum 
	entanglement is associated with  nonclassical correlations  between spatially separated quantum 
	systems. For a composite quantum system  the wavefunction cannot be written as a direct product of the wavefunctions of the subsystems. In Ref. 11  it was first shown that there exists a determinant connected with bipartite qubit and qutrit systems whose vanishing or non-vanishing values are related to existence of entanglement or non-entanglement.
	
	Considerable literature exists regarding various aspects of qubits, their superposition and applications in teleportation, cryptography and entanglement.$^{12-71}$ This work adds new insights  to the above body of literature  by 
    developing more insights regarding the determinant first derived in Ref. 11  to test entanglement for bipartite qubit and qutrit systems ; describing  a new teleportation protocol with enhanced security using non-maximally entangled states ; clarifying the physical implications of underlying SU(2) algebra for bipartite qubit systems and showing the existence of a SU(3) algebra for bipartite qutrit systems ; showing how time reversal 
    is linked to the occurrence of the above algebras; proposing plausible experimental scenarios in the context of two entangled electrons in the SU(2) case.
    
    The plan of the paper is : Section 2 is a quick review of basic entities and a recall  of certain fundamental results of Ref. 11. Section 3 describes a new protocol in teleportation with enhanced cryptographic security using non-maximally entangled states. Section 4, for the first time, describes exhaustively the underlying SU(2) symmetry for bipartite qubit systems. Section 5 is a new experimental proposal to physically clarify the underlying SU(2) symmetry. Section 6 contains our new results related to entanglement  for  bipartite qutrit systems. Section 7 gives a new perspective of  entanglement entropy for bipartite qubit and bipartite qutrit systems.Our conclusions are in  Section 8.
	\section{Bipartite Qubit states and  Entanglement$^{5-10, 11}$} 
	A general qubit is:
	$|\psi\rangle=\alpha |0\rangle+\beta|1\rangle$,  $|\alpha|^2$ + $|\beta|^2 = 1$ and  the computational basis is
	$|0\rangle=\begin{pmatrix}
		1 \\
		0
	\end{pmatrix}$ ,
	$|1\rangle=\begin{pmatrix}
		0 \\
		
		1
	\end{pmatrix}$.
	
	For a bipartite system , if  $\mathcal H_{A}$  and $\mathcal H_{B}$ are the Hilbert spaces of the two parts with 
	$\{|i\rangle_A\}$,
	$\{|\mu\rangle_B\}$ the orthonormal basis respectively, 
	then 
	an arbitrary bipartite state in $\mathcal H_A \otimes \mathcal H_B$ is
	$|\psi\rangle_{AB}= \Sigma_{\mu,i}    a_{i\mu}|i\rangle_A \otimes |\mu\rangle_B \equiv a_{i\mu}|i\rangle_A \otimes |\mu\rangle_B $ with summation over repeated indices.
	The density matrix is
	$\rho=|\psi\rangle\langle\psi|$ and 
	the expectation value of an operator $M$ is
	$\langle M\rangle=tr(M\rho)$.
	Further :
    $\langle M_A\rangle = _{AB}\langle\psi|M_A \otimes I_B|\psi\rangle_{AB}=tr(M_A \rho_A)$;
    $\rho_ A  = tr_{B}(|\psi\rangle_{AB}      
   {}_{AB}\langle\psi|)=tr_B(\rho_{AB}) 
	= _{B}\langle \mu|\rho_{AB}|\mu\rangle_{B} = a_{i\mu} a_{j\mu}^*|i\rangle_{A} {}_A \langle j|$;
     $\rho_B=tr_A(\rho_{AB})= {} _{A}\langle i|\rho_{AB}|i\rangle_{A}$ where 
	  $\rho_A, \rho_B$ are  the reduced density matrices. If $\mathcal H_{AB}\neq\mathcal H_A \otimes \mathcal H_B$, then the composite  state
	$|\psi\rangle_{AB}\neq|\psi\rangle_A\otimes|\psi\rangle_B$ and   $|\psi\rangle_{AB}$  is  said to be entangled.
	If $dim(\mathcal H_A)=dim(\mathcal H_B)=d$, then for maximal entanglement the reduced density matrices satisfy  $\rho_A=\rho_B=\frac{1}{d}I$ where 
	$I$ is the identity operator.
	
	The Bell States form the simplest example of maximally entangled states. They  also  form an orthonormal basis for a bipartite entangled state. The Bell States are
	\begin{equation}
    \label{bell}
		|\phi_0\rangle = \frac{|00\rangle+|11\rangle}{\sqrt{2}} ,
		\enskip
		|\phi_1\rangle=\frac{|01\rangle+|10\rangle}{\sqrt{2}} ,|\phi_{2}\rangle=\frac{|01\rangle-|10\rangle}{\sqrt{2}},   
		|\phi_3\rangle=\frac{|00\rangle-|11\rangle}{\sqrt{2}}
	\end{equation}
	with $\langle\phi_i|\phi_j\rangle=\delta_{ij}$ , $i$ and 
    $j$ takes values $0,1,2,3$.
	
	For the bipartite state $|\psi\rangle$  there exist orthonormal states $|i_A\rangle,|i_B\rangle$ for system $A$ and $B$ respectively, such that
	\begin{equation}
    \label{smdgen}
		|\psi\rangle=\sum_i\sqrt{\mu_i}|i_A\rangle|i_B\rangle
	\end{equation}
	where $\mu_i$ are non-negative real numbers such that,
	$\sum_i\mu_i=1$. This is the Schmidt decomposition$^{4}$. Schmidt number is the number of non-zero eigen values of the reduced density matrix. 
	$\rho_A$ and $\rho_B$ have the  same eigenvalues $\mu_i$. If the number of terms in equation (\ref{smdgen}) is greater than one  then the state $|\psi\rangle$ is entangled, if equal to  unity then $|\psi\rangle$ is not entangled.

	Consider the 2-state systems A and B  with Hilbert spaces $\mathcal H_A$ and $\mathcal H_B$ and basis vectors 
	$(|0\rangle_A , |1\rangle_A)$ 
	and 
	$(|0\rangle_B ,|1\rangle_B)$respectively.
	The basis vectors for the bipartite state $|\psi_{AB}\rangle$ are:
	$|00\rangle,|01\rangle,|10\rangle,|11\rangle$. The first entry in the ket denotes system A , second entry denotes system B  and
	$\langle ij|i'j'\rangle=\delta_{i,i'}\delta_{j,j'}$
	where $(i,j)$ is $0$ or $1$.
	The most general bipartite state can be written as$^{11}$:
	\begin{equation}
    \label{state}
|\psi\rangle=a_{00}|00\rangle+a_{01}|01\rangle+a_{10}|10\rangle+a_{11}|11\rangle 
	\end{equation}
with $ a_{00}^2+a_{01}^2+a_{10}^2+a_{11}^2=1$.
	Consider the matrix $A$ 
    (for simplicity  $a_{ij}$ is real)
	\begin{equation}
    \label{A}
		A=\begin{pmatrix}
			a_{00} && a_{01}\\ 
			
			a_{10} && a_{11}
		\end{pmatrix}
	\end{equation}
	The density matrix is 
	$\rho=|\psi\rangle\langle\psi|$ and 
	reduced density matrix is 
	\begin{equation} 
    \label{reduce}
		\rho_A = Tr_B(\rho) 
		=\begin{pmatrix}
			a_{00}^2+a_{01}^2 && a_{00}a_{10}+a_{01}a_{11}\\
			
			a_{00}a_{10}+a_{01}a_{11} && a_{11}^2+a_{10}^2
		\end{pmatrix}=AA^{\dagger}
	\end{equation}
	The two eigen values are $\mu_{i}$,(i=1,2) :
	\begin{equation}
 \begin{split}
 \label{E1}
     \mu_i =(1/2)(1 \mp \sqrt{1-4(DetA)^2}) \enskip; \enskip for\enskip i=1,2
 \end{split}	
	\end{equation}
The Schmidt decomposition is
	\begin{equation}
		\label{S.D.S}
|\psi\rangle=\sum_{i=1}^2\sqrt{\mu_i}|\rangle|\mu_i\rangle
	\end{equation}
	Clearly if $DetA =0$ then, $=0$,  $\mu_2 \neq 0$, and
	\begin{equation}
		|\psi\rangle=\sqrt{\mu_2}|\mu_2\rangle|\mu_2\rangle
	\end{equation}
	 and so the Schmidt number is unity and state is {\it not} entangled.Thus the condition for entanglement in a general bipartite basis is $DetA \neq 0$. 
		
 $|\psi\rangle$ is maximally entangled if
\begin{equation}
		\rho_A=
		\begin{pmatrix}
			a_{00}^2+a_{01}^2 && a_{00}a_{10}+a_{01}a_{11}\\
			
			a_{00}a_{10}+a_{01}a_{11} && a_{11}^2+a_{10}^2 \\
		\end{pmatrix}
		=\frac{I}{2}
	\end{equation}
	So the range of $DetA$ is
	\begin{equation}
		0 \leq|DetA|\leq \frac{1}{2}
		\label{range}
	\end{equation}
	The same results hold for $\rho_B$.
	Thus for maximal violation $0<|DetA|< \frac{1}{2}$ and greater the value of $|DetA|$ the more entangled the state is. As an example, consider the situation discussed in reference$^{45}$. The state is $|\psi\rangle = cos\theta |00\rangle + sin\theta |11\rangle$ and  here $DetA=sin (2\theta)/2$. When  
	$DetA = sin (2\theta)/2 = 1/2$  the state is  maximally entangled for  $\theta=n\pi/4$ and not entangled for $DetA=sin (2\theta)/2=0$, that is for $\theta=n\pi/2$. 

We continue the above discussion in the Bell basis. A general  state $|\eta\rangle$ is
\begin{equation}
\label{eta in bell basis}
\begin{split}
|\eta\rangle = b_0|\phi_0\rangle + b_1|\phi_1\rangle + b_2|\phi_2\rangle + b_3|\phi_3\rangle \\ 
= (1/2)^{1/2} [ (b_0+b_3)|00\rangle 
		+ (b_1+b_2)|01\rangle\\
		+  (b_1 - b_2)|10\rangle +(b_0-b_3)|11\rangle] 
\end{split}
\end{equation}
	where, $\langle\eta|\eta\rangle=1$ ,
	i.e. $b_0^2+b_1^2+b_2^2+b_3^2=1$, where we have used equation (\ref{bell}).

	Identifying
	$a_{00}= (1/2)^{1/2}(b_0+b_3) , a_{01}= (1/2)^{1/2}(b_1+b_2)  ,   
	a_{10}= (1/2)^{1/2}(b_1-b_2) , a_{11}= (1/2)^{1/2}(b_0-b_3)$  , 
	we get back equation (\ref{state}) and $|\eta\rangle \equiv |\psi\rangle $. The normalization conditions are also  consistent with each other.
	
	For the Schmidt decomposition of $|\eta\rangle$
	define the matrix C as
	\begin{equation}
		C= \frac{1}{\sqrt{2}}\begin{pmatrix}
			b_0+b_3 && b_1+b_2\\
			b_1-b_2 && b_0-b_3
		\end{pmatrix}
	\end{equation}
	Here $b_i$'s are real. The density matrix is $\rho=|\eta\rangle\langle\eta|$ and the reduced density matrix for subsystem A is 
	\begin{equation}
	\rho_A=tr_B(|\eta\rangle\langle\eta|)= _B\langle0|\eta\rangle\langle\eta|0\rangle_B +  _B\langle1|\eta\rangle\langle\eta|1\rangle_B 
    \end{equation}
The two eigenvalues of  $\rho_A$ are (i=1,2):
\begin{equation}
\label{belleigen}
\mu_i = (1/2)(1 \mp\sqrt{1 -(DetC)^2})
\end{equation}
The Schmidt decomposition series will then be:

  $|\eta\rangle=\sum_{i=1}^2\sqrt{\mu_i}|\mu_i\rangle$

But if $DetC=0$ then $=0$ and the Schmidt decomposition series has only one term :$|\eta\rangle=\sqrt{\mu_2}|\mu_2\rangle|\mu_2\rangle$
and the state is not entangled. Thus the condition  for not entanglement in the Bell basis is $ DetC=0$ which means $b_0^2 + b_2^2 = b_1^2+b_3^2$, i.e. $\langle\phi_0|\eta\rangle^2+\langle\phi_2|\eta\rangle^2=\langle\phi_1|\eta\rangle^2+\langle\phi_3|\eta\rangle^2$. 
 Thus ,  for every basis system matrices exist whose determinant being zero means non-entanglement while a non-zero value implies entanglement.

\section{Teleportation using non-maximally entangled states:}
\begin{figure}
\centering
\begin{quantikz} 
    \lstick{\(|\psi\rangle_I=\alpha|0\rangle+\beta|1\rangle\)} & \ctrl{1} &\qw  & \gate{H} & \qw & \qw  \\
    \qw & \targ{} & \qw & \qw & \meter{} & \qw \\
     \lstick{\(|\gamma\rangle=a_{00}|00\rangle+a_{11}|11\rangle\)} \Bigg{\{}\\
     \qw & \qw & \qw & \qw &\meter{} &  \qw   
\end{quantikz} \\
\caption{Teleportation Circuit}
\end{figure}

Let the information qubit be $|\psi\rangle_I=\alpha|0\rangle+\beta|1\rangle$ ,($\alpha^2+\beta^2=1$). To send the information qubit via a  quantum teleportation channel we use the {\it non-maximally entangled state}  $|\gamma\rangle=a_{00}|00\rangle+a_{11}|11\rangle$ ,
($a_{00}^2+a_{11}^2=1$).
This is a non-maximally entangled state as the probability amplitudes $a_{00}\neq 1/\sqrt {2}$ , $a_{11}\neq 1/\sqrt{2}$ and $DetA=a_{00}.a_{11}\neq 1/2$ (from equation \ref{range}). Consider the teleportation protocol Fig. 1:
            \begin{equation}
            \label{teleportation}
                \begin{split}
                    |\psi_0\rangle=|\psi\rangle_I \otimes|\gamma\rangle=(\alpha|0\rangle +\beta|1\rangle) \otimes (a_{00}|00\rangle+a_{11}|11\rangle)\\\xrightarrow[]{CNOT} [ \alpha|0\rangle \otimes (a_{00}|00\rangle+a_{11}|11\rangle) ]+[\beta|1\rangle \otimes (a_{00}|10\rangle+a_{11}|01\rangle) ]
                  \\  \xrightarrow[]{HADAMARD} |\psi\rangle_f=\frac{1}{\sqrt{2}}[|00\rangle(\alpha a_{00}|0\rangle+\beta a_{11}|1\rangle) +|11\rangle(\alpha a_{11}|1\rangle-\beta a_{00}|0\rangle) \\
                    +|01\rangle(\alpha a_{11}|1\rangle+\beta a_{00}|0\rangle)
                    +|10\rangle(\alpha a_{00}|0\rangle-\beta a_{11}|1\rangle)]
                \end{split}
            \end{equation}
          
          Now if Alice measures in basis $|00\rangle$, then Bob's measurement will be, \begin{equation}
          \label{measure}
            \begin{split}
                M_0=|\langle 0| \psi\rangle_f|^2=\alpha^2 a_{00}^2/2 \enskip; \enskip M_1=|\langle 1 | \psi\rangle_f|^2=\beta^2 a_{11}^2/2 \\ and \enskip M_0 M_1=\alpha^2 \beta^2 |DetA|^2/4
                \end{split}
            \end{equation}
			where $M_0$ and $M_1$ are Bob's measurements in $|0\rangle$ and $|1\rangle$ basis respectively. $M_0M_1$ does not change even if Alice makes her measurement in any of the other three basis \{$|01\rangle,|10\rangle,|11\rangle$\}
        The fidelity of the circuit is
            \begin{equation}
                \label{fidelity}
                F= _I\langle \psi|\rho_f|\psi\rangle_I=(a_{00}\alpha+a_{11}\beta)^2
                = [a_{00}\alpha + \sqrt{1- a_{00}^2}\enskip \beta]^2
            \end{equation}
            where $ \rho_f= |\psi\rangle_f {}_f \langle \psi|$.
        For the fidelity to be an extremum;
              $  \frac{dF}{da_{00}}=0$ 
            i.e. $a_{00}=\pm \alpha ;\pm \beta$, so the two extremal fidelities ($F_1;F_2$) are 
            \begin{equation}
            \label{F1&F2}
                F_1=4|DetA|^2 \enskip for \enskip a_{00}=\pm \beta \enskip \text{and} \enskip F_2=1 \enskip for \enskip a_{00}=\pm \alpha
          \end{equation}The second order derivative of F i.e $\frac{d^2F}{da_{00}^2}=2\frac{(2|\alpha|^2-1)^2}{|\alpha|^2}>0$ for $a_{00}=\pm \beta$ and $\frac{d^2F}{da_{00}^2}=\frac{-2}{|\beta|^2} <0$ when $a_{00}=\pm \alpha$, since $|\alpha|^2$ and $|\beta|^2$ is always positive. So $F_1$ is the minimum fidelity and $F_2$ is the maximum fidelity. Thus we show that teleportation using non-maximally entangled state is possible in our formalism. In the usual case Alice has to tell Bob through classical communication channel in which of the basis \{$| 00\rangle,|01\rangle,|10\rangle,|11\rangle$\}  she has made her own measurement and accordingly Bob makes his measurements. In our protocol additionally $|DetA|$ must also be given to Bob along with the basis in which Alice has made her measurement. Then using  $\alpha^2+\beta^2=1$ and equation (\ref{measure}) Bob can calculate $\alpha$ and $\beta$. More over the fidelity lies between $F_1$and $F_2$. For the maximally entangled state $|DetA|^2=1/4$ and  we get F=1 (which is an already known result).
 \section{ The underlying SU(2) algebra for bipartite qubit systems  :}
Now we construct the $A$ matrix for  the four Bell States and show that there exists an underlying $SU(2)$ algebra. 
$|\psi\rangle=a_{00}|00\rangle+a_{01}|01\rangle+a_{10}|10\rangle+a_{11}|11\rangle$ and the $A$ matrix for this state is
	$$A=\begin{pmatrix}
		a_{00} && a_{01}\\
		
		 a_{10} && a_{11}
		\end{pmatrix}$$
		When
	 $|\psi\rangle= |\phi_0 \rangle$;
	$a_{00}=\langle 00|\phi_0 \rangle=1/\sqrt{2}$,
	$a_{01}=\langle 01|\phi_0 \rangle=0$, 
	$a_{10}=\langle 10|\phi_0 \rangle=0$, 
	$a_{11}=\langle 11|\phi_0 \rangle=1/\sqrt{2}$
	so that
	\begin{equation}
    \label{A0}
		A_0=\frac{1}{\sqrt{2}}\begin{pmatrix}
			1 && 0\\
			
			 0 &&1
\end{pmatrix}			 
		=\frac{I}{\sqrt{2}}
	\end{equation}
	
	Similarly, for $|\psi\rangle= |\phi_1 \rangle ,
    |\phi_2 \rangle, |\phi_3 \rangle$, the relevant matrices are:
	\begin{equation}
    \label{A1-3}
		A_1=\frac{1}{\sqrt{2}}\begin{pmatrix}
			0&&1\\
			
			 1&&0 \end{pmatrix}
		=\frac{1}{\sqrt{2}}\sigma_1 ;
        A_2 = \frac{1}{\sqrt{2}}\begin{pmatrix}
			0&&1\\
			
			-1&&0
			\end{pmatrix}
		=\frac{i}{\sqrt{2}}\sigma_2 ; 
A_3=\frac{1}{\sqrt{2}}\begin{pmatrix}
			1&&0\\
			
			0&&-1
			\end{pmatrix}
		=\frac{1}{\sqrt{2}}\sigma_3
\end{equation}
		where $I$ is the unit matrix and $\sigma_i$
	are Pauli Matrices.
	
  Commutation  and anticommutation relations satisfied by  these matrices are :
	\begin{equation}
	     \label{SU2-d}
        [A_i,A_j]=\sqrt{2} \sum_k \epsilon'_{ijk}A_k \quad ; \{A_i,A_j\}=(-1)^{\frac{(i^3+j^3)-(i+j)}{4}}\delta_{ij}
    \end{equation}
    with  $\epsilon'_{ijk}=\theta(i,j)\epsilon_{ijk}$, for i and j=1,2,3
    \begin{table}[H]
    \centering
    \renewcommand{\arraystretch}{0.5} 
    \begin{tabular}{|c|c|c|}
        \hline
        \textbf{ijk} & \textbf{$\epsilon_{ijk}$} & \textbf{\(\theta(i,j)\)} \\ 
        \hline
        123 & \(\epsilon_{123}\)=1 & \(\theta(1,2)=-1\) \\ 
            &            & \(\theta(2,3)=-1\) \\ 
            &            & \(\theta(3,1)=+1\) \\ 
      \hline
    \end{tabular}
    \caption{Table for $\theta(i,j)$ }
     \label{theta(ij)qubit}
\end{table}
 Note that $\theta(i,j)=(-1)^{i+j}$.
	
    Now, $\sigma_2=\sigma_2^{\dagger}$. Therefore 
	$A_2 = -A_2^{\dagger}$. Thus $A_2$ is pure imaginary.
	So one may write $A_2\equiv K^{\dagger}U$, 
	where $K^{\dagger}$ is complex conjugation and $U$ is a unitary operator. Similar scenarios have  been reported elsewhere$^{29}$ in a totally different context.
  
    Redefining variables :
	$A_1'=\sqrt{2}A_1$ , $A_2'=\sqrt{2}e^{(-i\pi)/2}A_2$, $A_3'=\sqrt{2}A_3$, 
	one obtains the $SU(2)$ algebra
	\begin{equation}
		\label{SU2}
		[A_i',A_j']=2i\epsilon_{ijk}A_k' \enskip; \enskip
        \{A_i ' , A_j ' \}=2\delta_{ij}
	\end{equation}

We now  define four 4-dimensional matrices :
\begin{equation}
\label{tau-dash def}
\tau_0' = \begin{pmatrix} 0 & A_0' \\ A_0' & 0 \end{pmatrix} \enskip, \enskip \tau_i' = \begin{pmatrix} A'_i & 0 \\ 0 & -A'_i \end{pmatrix} \enskip for \enskip i=1,2,3
\end{equation}
where the $\tau_i'$ (for i=0,1,2,3) are hermitian and unitary.The actions of the matrices 
$\tau_i'$ on the basis set $|\phi_i\rangle,  i=0,1,2,3$ are               

\begin{subequations}
\label{tau_operators}
\begin{align}
    \tau_0'|\phi_0\rangle = +1|\phi_1\rangle\enskip,\enskip
    \tau_0'|\phi_1\rangle = +1|\phi_0\rangle \enskip, \enskip 
\tau_0'|\phi_2\rangle=+1|\phi_3\rangle\enskip,\enskip
\tau_0'|\phi_3\rangle = +1|\phi_2\rangle, \label{tau0'} \\ 
    \tau_1'|\phi_0\rangle = +1|\phi_2\rangle\enskip,\enskip
    \tau_1'|\phi_1\rangle = +1|\phi_3\rangle\enskip,\enskip  
    \tau_1'|\phi_2\rangle = +1|\phi_0\rangle\enskip,\enskip \tau_1'|\phi_3\rangle = +1|\phi_1\rangle, \label{tau1'} \\ 
    \tau_2'|\phi_0\rangle = +i|\phi_1\rangle\enskip,\enskip \tau_2'|\phi_1\rangle = -i|\phi_0\rangle\enskip,\enskip  
    \tau_2'|\phi_2\rangle = -i|\phi_3\rangle\enskip,\enskip \tau_2'|\phi_3\rangle = +i|\phi_2\rangle, \label{tau2'} \\ 
    \tau_3'|\phi_0\rangle = +1|\phi_0\rangle\enskip,\enskip \tau_3'|\phi_1\rangle = -1|\phi_1\rangle\enskip,\enskip  
    \tau_3'|\phi_2\rangle = -1|\phi_2\rangle\enskip,\enskip
    \tau_3'|\phi_3\rangle = +1|\phi_3\rangle. \label{tau3'}
\end{align}
\end{subequations}
  $\tau_3'$ is the diagonal operator with eigenvalues +1,+1 ; -1,-1. Only two are distinct,{\it viz.}, +1   and  -1. As in the electron spin case, there are only two internal  degrees of freedom. {\it Thus,  the increase in number of basis vectors to four for a bipartite entangled qubit state does not increase the number of effective internal degrees of freedom.} The +1 eigenvalue correspond to the eigen basis states  $|\phi_0\rangle$ and $|\phi_3\rangle$ while the  -1 eigenvalue corresponds to $|\phi_1\rangle$ and $|\phi_2\rangle$. In case of a single electron (2-dim internal space) the basis $|0\rangle$ and $|1\rangle$ corresponds to different internal degrees of freedom , whereas here (4-dim internal space) two eigen basis have +1 and the other two -1 eigenvalues; {\it i.e., there is no increase in the internal degrees of freedom for two entangled electrons}.  However the analogues of operations of $\sigma_1,\sigma_2,\sigma_3$ on $|0\rangle$ and $|1\rangle$ exist here also for $\tau_i'$ matrices operating on $|\phi_i\rangle's$ (i=0,1,2,3). This is verifiable from equations (\ref{tau0'}) to equation (\ref{tau3'}). As in the case of unentangled electron states where $\sigma_1$ used to flip the spin ; here there are two  flip operators $\tau_0'$ and $\tau_1'$ which takes a basis corresponding to the +1 eigenvalue ($|\phi_0\rangle$ or $|\phi_3\rangle$) to another basis corresponding to the -1 eigenvalue ($|\phi_1\rangle$ or $|\phi_2\rangle$).

In case of a single electron the internal space is 2-dimensional. For measuring the spin we align one of the internal basis axis with the  z-axis of the external physical  space. Note that the projective measurement of a 2-dimensional  vector in internal space of single qubit gives us a 1-dimensional  line along the aligned basis. In case of bipartite qubit entangled space the internal space is 4-dimensional. A projection is again a 1-dimensional line in 3-dimensions of the internal space. In order to make a measurement we have to ensure that this 1-dimensional (internal space) line is aligned with the measurement axis of the 3-dimensional measurement axis of the real physical space.

The repetitive operation of $\tau_i'$ (i=0,1,2) matrices on the Bell basis are :

  \begin{align*}
|\phi_0\rangle &\xrightarrow{\tau_0'} |\phi_1\rangle \xrightarrow{\tau_0'} |\phi_0\rangle 
\xrightarrow{\tau_0'} |\phi_1\rangle \xrightarrow{\tau_0'} |\phi_0\rangle  
& |\phi_1\rangle &\xrightarrow{\tau_0'} |\phi_0\rangle \xrightarrow{\tau_0'} |\phi_1\rangle 
\xrightarrow{\tau_0'} |\phi_0\rangle \xrightarrow{\tau_0'} |\phi_1\rangle  \\
|\phi_0\rangle &\xrightarrow{\tau_1'} |\phi_2\rangle \xrightarrow{\tau_1'} |\phi_0\rangle 
\xrightarrow{\tau_1'} |\phi_2\rangle \xrightarrow{\tau_1'} |\phi_0\rangle  
& |\phi_1\rangle &\xrightarrow{\tau_1'} |\phi_3\rangle \xrightarrow{\tau_1'} |\phi_1\rangle 
\xrightarrow{\tau_1'} |\phi_3\rangle \xrightarrow{\tau_1'} |\phi_1\rangle  \\ 
|\phi_0\rangle &\xrightarrow{\tau_2'} i|\phi_1\rangle \xrightarrow{\tau_2'} |\phi_0\rangle  
\xrightarrow{\tau_2'} i|\phi_1\rangle \xrightarrow{\tau_2'} |\phi_0\rangle  
& |\phi_1\rangle &\xrightarrow{\tau_2'} -i|\phi_0\rangle \xrightarrow{\tau_2'} |\phi_1\rangle  
\xrightarrow{\tau_2'} -i|\phi_0\rangle \xrightarrow{\tau_2'} |\phi_1\rangle  \\[10pt]
|\phi_2\rangle &\xrightarrow{\tau_0'} |\phi_3\rangle \xrightarrow{\tau_0'} |\phi_2\rangle 
\xrightarrow{\tau_0'} |\phi_3\rangle \xrightarrow{\tau_0'} |\phi_2\rangle  
& |\phi_3\rangle &\xrightarrow{\tau_0'} |\phi_2\rangle \xrightarrow{\tau_0'} |\phi_3\rangle 
\xrightarrow{\tau_0'} |\phi_2\rangle \xrightarrow{\tau_0'} |\phi_3\rangle  \\
|\phi_2\rangle &\xrightarrow{\tau_1'} |\phi_0\rangle \xrightarrow{\tau_1'} |\phi_2\rangle 
\xrightarrow{\tau_1'} |\phi_0\rangle \xrightarrow{\tau_1'} |\phi_2\rangle  
& |\phi_3\rangle &\xrightarrow{\tau_1'} |\phi_1\rangle \xrightarrow{\tau_1'} |\phi_3\rangle 
\xrightarrow{\tau_1'} |\phi_1\rangle \xrightarrow{\tau_1'} |\phi_3\rangle  \\ 
|\phi_2\rangle &\xrightarrow{\tau_2'} -i|\phi_3\rangle \xrightarrow{\tau_2'} |\phi_2\rangle  
\xrightarrow{\tau_2'} -i|\phi_3\rangle \xrightarrow{\tau_2'} |\phi_2\rangle  
& |\phi_3\rangle &\xrightarrow{\tau_2'} i|\phi_2\rangle \xrightarrow{\tau_2'} |\phi_3\rangle  
\xrightarrow{\tau_2'} i|\phi_2\rangle \xrightarrow{\tau_2'} |\phi_3\rangle  
\end{align*}
Note that the $\tau_i'$ satisfy a Clifford algebra $(i, j =0,1,2,3)$:
\begin{equation}
\label{tau-dash anti}
   \{\tau_i',\tau_j'\}= \tau_{i}'\tau_{j}'+ \tau_{j}'\tau_{i}'=2\delta_{ij}
\end{equation} 
 
\section{A suggested experiment:}
We now clarify the  origin of this SU(2) algebra for entangled states based on  $|\phi_0\rangle$ and $|\phi_1\rangle$. Exactly similar arguments hold for $|\phi_2\rangle$ and $|\phi_3\rangle$. Note $|\phi_i\rangle$'s are eigenstates of $\tau_3'$ as shown previously.
We propose an experimental set up to pin point the physical picture related to the $A$ matrices. This proposal is an analogue of the Stern-Gerlach(S.G) set up. We use an entangled pair of electrons in an  inhomogenous magnetic field. The steps are :

(1) Consider the schematic diagram Fig.3. The Bell state 
$|\phi_0\rangle$ is first projected into the bipartite basis: ($|00\rangle\langle00|\phi_0\rangle=\frac{1}{\sqrt{2}}\begin{pmatrix}
    1 & 0 & 0 & 0\end{pmatrix}^T$
 or $|11\rangle\langle11|\phi_0\rangle=\frac{1}{\sqrt{2}}\begin{pmatrix}
    0 & 0 & 0 & 1\end{pmatrix}^T$).
    
(2)Now we send each of the bipartite states through a blackbox which reduces the 4-dim. space into a direct product of two 2-dim subspaces i.e $|00\rangle = |0\rangle |0\rangle$ and 
$|11\rangle = |1\rangle |1\rangle$. 
(This means :$\begin{pmatrix}
    1 & 0 & 0 & 0\end{pmatrix}^T \xrightarrow{Blackbox}\begin{pmatrix}
    1 & 0\end{pmatrix}^T \otimes \begin{pmatrix} 1 & 0\end{pmatrix}^T$and $\begin{pmatrix}
    0& 0 & 0 & 1\end{pmatrix}^T \xrightarrow{Blackbox}\begin{pmatrix}
    0 & 1\end{pmatrix}^T \otimes \begin{pmatrix}
    0 & 1\end{pmatrix}^T$). 
    $|0\rangle$ and $|1\rangle$ are the eigenstates of $\sigma_3$. These are  then passed through a region analogous to Stern-Gerlach and follow the trajectories shown in the figures Fig.4-Fig.7  and register themselves on the blue screen. 
    Thus we have the usual Stern-Gerlach scenario.

(3) We follow exactly the same procedures for 
$|\phi_1\rangle $ as described above in steps 1 and 2. 
The results are summarized in Fig.7- Fig.9.

We have thus given a physical interpretation of the underlying SU(2) algebra in our formalism  associated with an entangled pair of electrons.    
\begin{figure}
    \centering
        \includegraphics[width=1.0\linewidth]{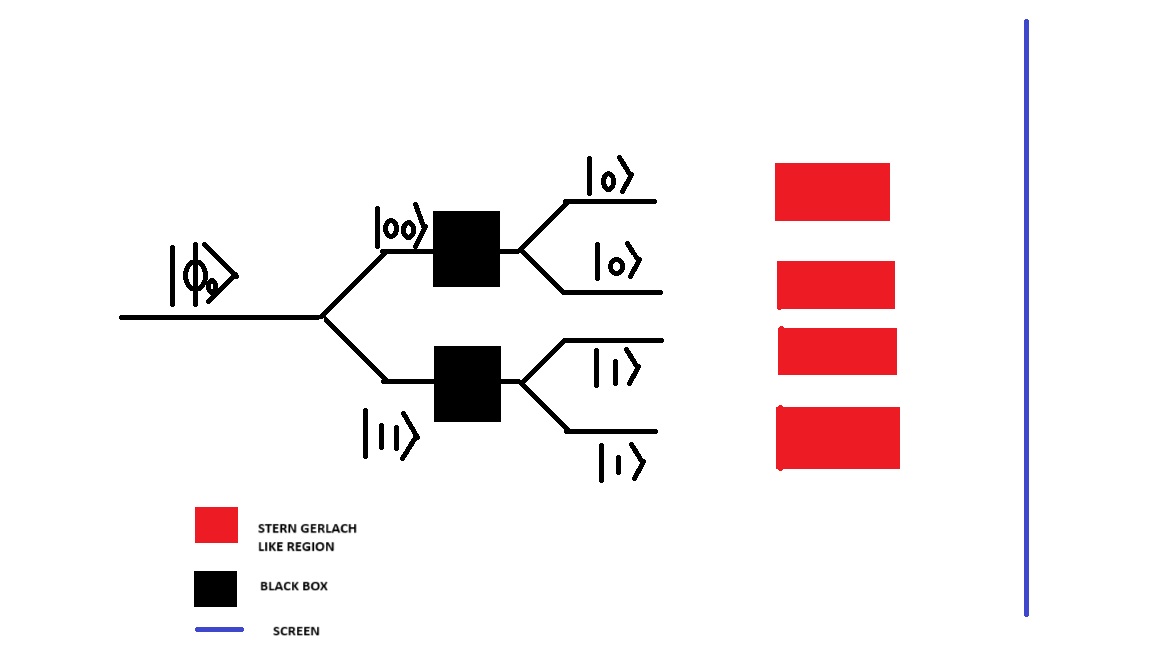} 
       \caption{Block Diagram for experiment on $|\phi_0\rangle$} 
\end{figure}
    \begin{figure}
           \centering
           \includegraphics[width=0.5\linewidth]{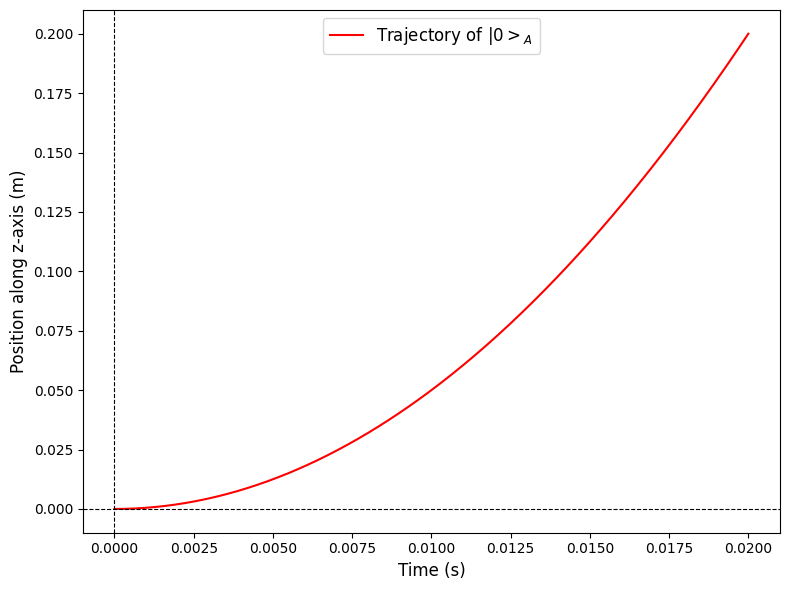}
           \caption{The Trajectory of $|0\rangle_A$ in S.G like region}
           \label{fig:enter-label}
       \end{figure}   
   \begin{figure}
       \centering
       \includegraphics[width=0.5\linewidth]{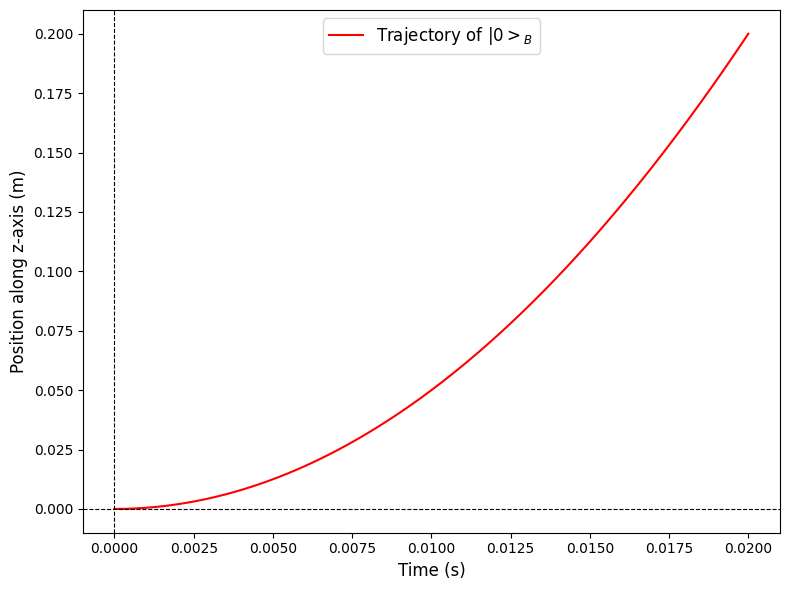}
       \caption{The Trajectory of $|0\rangle_B$ in S.G like region}
       \label{fig:enter-label}
   \end{figure}
\begin{figure}
    \centering
    \includegraphics[width=0.5\linewidth]{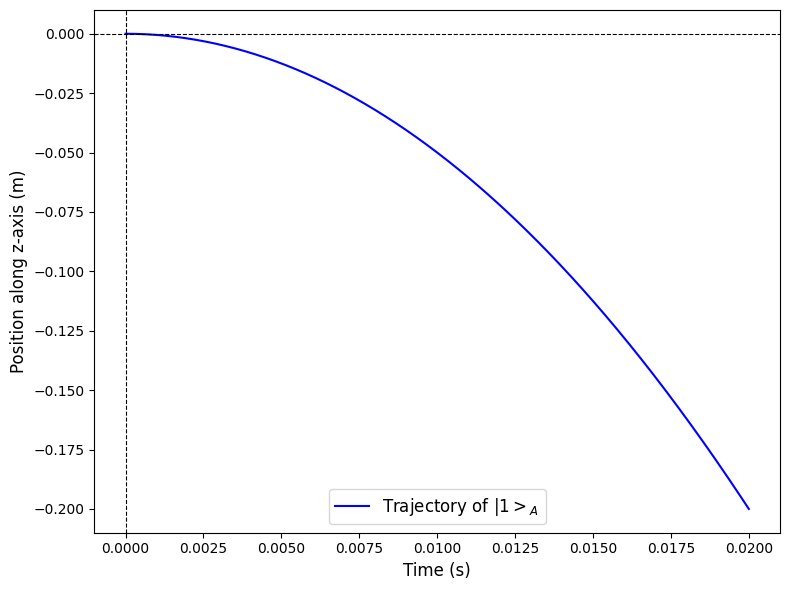}
    \caption{The Trajectory of $|1\rangle_A$ in S.G like region}
    \label{fig:enter-label}
\end{figure}
\begin{figure}
    \centering
    \includegraphics[width=0.5\linewidth]{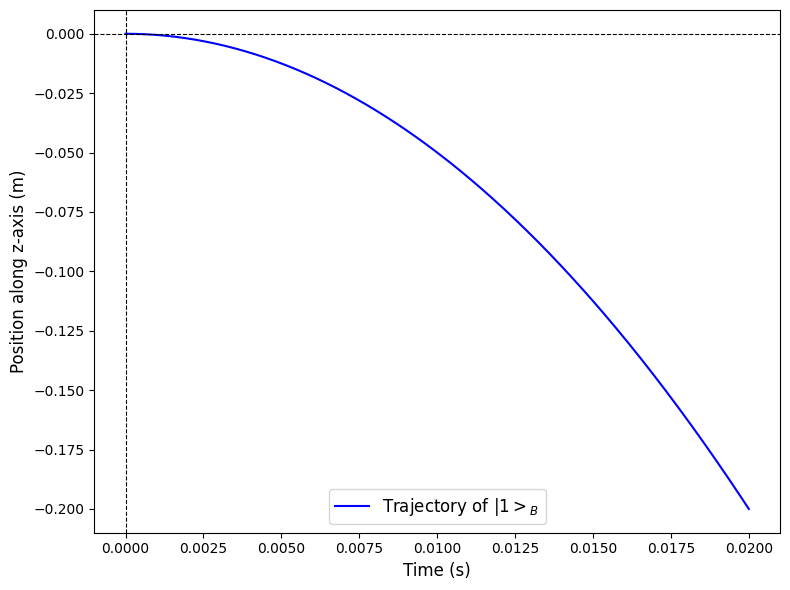}
    \caption{The Trajectory of $|1\rangle_B$ in S.G like region}
    \label{fig:enter-label}
\end{figure}
\begin{figure}
      \centering
      \includegraphics[width=1.0\linewidth]{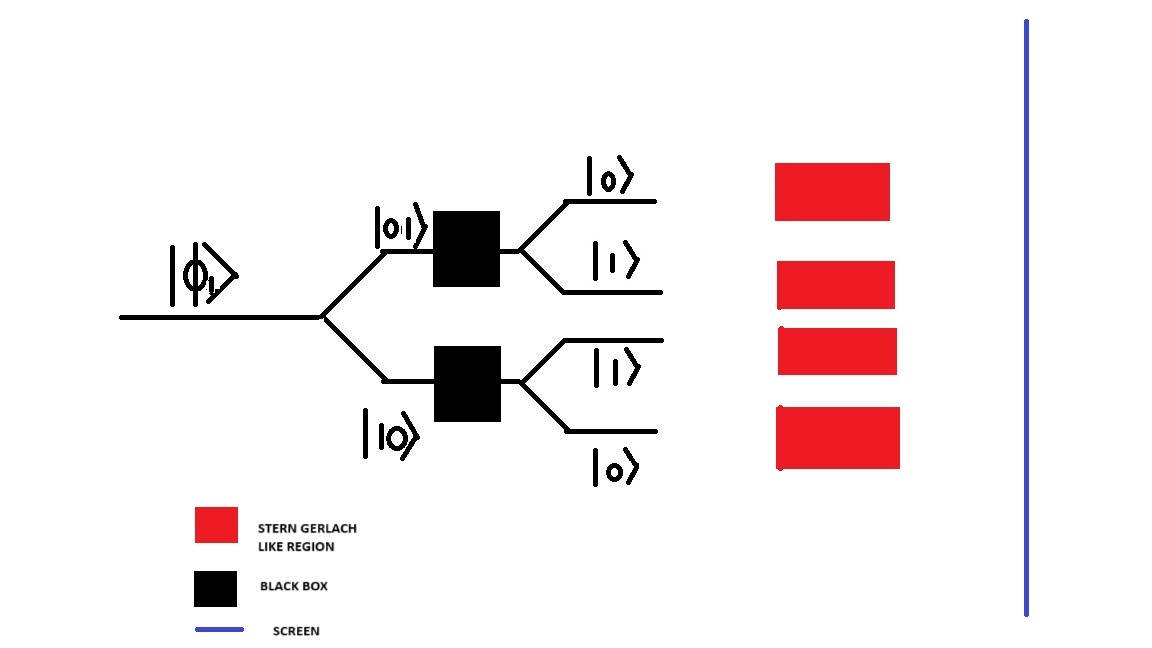}
      \caption{Block Diagram for experiment on $|\phi_1\rangle$}
      \label{blockphi1}
  \end{figure}
  \begin{figure}
      \centering
      \includegraphics[width=0.5\linewidth]{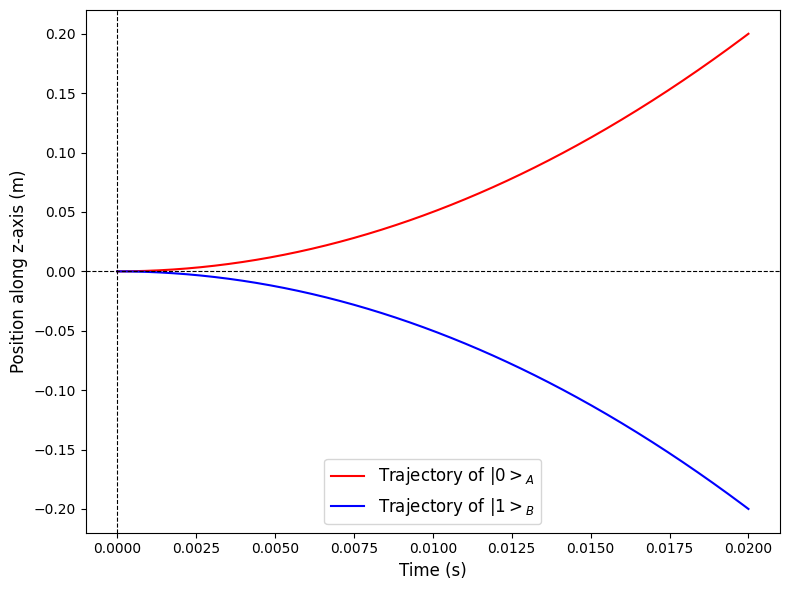}
      \caption{Trajectory of $|01\rangle $ in S.G like region}
      \label{fig:enter-label}
  \end{figure}
\begin{figure}
    \centering
    \includegraphics[width=0.5\linewidth]{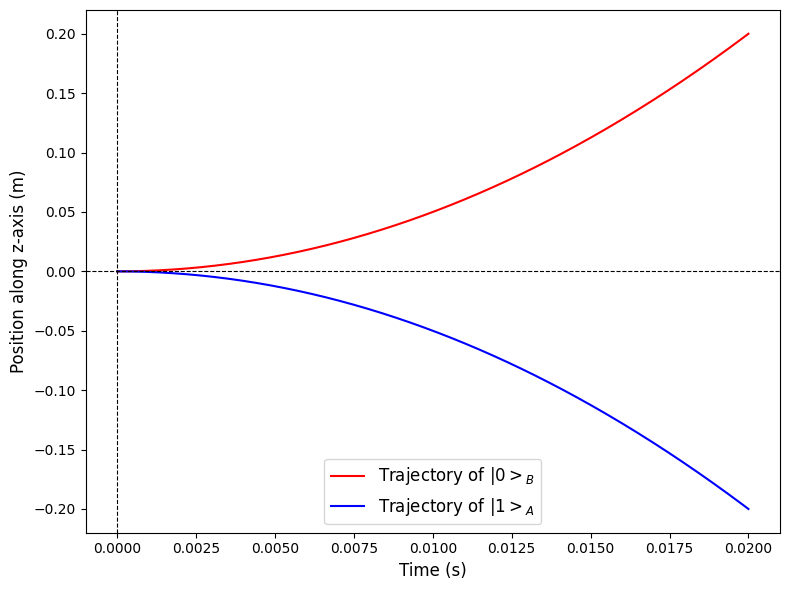}
    \caption{Trajectory of $|10\rangle $ in S.G like region}
    \label{fig:enter-label}
\end{figure}
  
    \section{The underlying SU(3) algebra for bipartite qutrit systems$^{11}$}
	A natural generalisation of two state systems to a three state system is the qutrit. We now study entanglement in a bipartite qutrit system$^{75-77}$.The most general qutrit is $|\mu\rangle=\alpha|0\rangle+\beta|1\rangle+\gamma|2\rangle$, where
	the column vectors 
	$|0\rangle=\begin{pmatrix}
		1 && 0 && 0
		\end{pmatrix}^T\enskip $\ ;
	$|1\rangle=\begin{pmatrix}
		0 && 1 && 0
	\end{pmatrix}^T\enskip $;
	$|2\rangle=\begin{pmatrix}
		0 && 0 && 1
	\end{pmatrix}^T\enskip $
	 form an orthonormal basis.
	Let there be two systems A and B belonging to 3-dimensional Hilbert spaces $\mathcal H_{A}$ and  $\mathcal H_{B}$ respectively. The basis vectors for the composite system AB belonging to the 9-dimensional Hilbert space, $\mathcal{H_{AB}}$ are 
  $\{ |00\rangle , |01\rangle , |02\rangle , |10\rangle , |11\rangle , |12\rangle , |20\rangle , |21\rangle , |22\rangle \}$ , 
  where $\langle i i'| j j'\rangle = \delta_{ij}\delta_{i'j'}$  with $i,j=0,1,2$. A general quantum state in the composite system AB is:
\begin{equation}
\label{chi}
|\chi\rangle_{AB}  
= \sum_{i=0}^2\sum_{j=0}^2a_{ij}|ij\rangle 
\end{equation}
Define the matrix,
	\begin{equation}
    \label{P}
		P=\begin{pmatrix}
			a_{00} && a_{01} && a_{02}\\
			
			 a_{10} && a_{11} && a_{12}\\
			  a_{20} && a_{21} && a_{22}
			\end{pmatrix}
	\end{equation}
	The density matrix for the state $|\chi \rangle_{AB}$ is
	$\rho_{AB}$ and 
	the reduced density matrix is
	$\rho_A= PP^{\dagger}$. 
	 $P$ is a 3x3 matrix and it will have three eigenvalues $\mu_0, \mu_1, \mu_2$. Then the eigenvalues of $\rho_A=PP^{\dagger}$ are $|\mu_0|^2,|\mu_1|^2$ and $|\mu_2|^2$. $|Det\rho_A|=|DetP|^2=|\mu_0|^2.|\mu_1|^2.|\mu_2|^2 $. Thus we get $|DetP|=|\mu_0|\mu_1||\mu_2|$. Now $TrP=\mu_0+\mu_1+\mu_2$ . From  the normalization of the density matrix , we get the trace $Tr\rho_A=|\mu_0|^2+|\mu_1|^2+|\mu_2|^2 =1$
	The Schmidt decomposition series for $\rho_A$ is
	\begin{equation}
		\label{qutrit schmidt}
|\chi\rangle_{AB}=\sum_{k=0}^2\mu_k|\mu_k\rangle|\mu_k\rangle
	\end{equation}
	We get the exact  analytical solutions for the eigenvalues $\mu_k$  by diagonalizing the P matrix, and writing them in terms of trace of P( sum of eigenvalues) and determinant of P (product of eigenvalues), where $\mu_k$'s are given by the roots of the cubic equation :
    
    $\mu^3 - TrP\mu^2 + \frac{2(TrP)^2 - 1}{2} \mu - Det P = 0$

    The roots are :
    
    $\mu_k = \frac{TrP}{3} + \frac{\omega^k C + \frac{\Delta_0}{\omega^k C}}{3}, \quad k = 0,1,2.$ where $ \omega = e^{2\pi i /3} $ (cube root of unity)
$ \Delta_0 = \frac{-(TrP)^2+3}{2}$ ; 
$ \Delta_1 = \frac{5(TrP)^2-9TrP-54|DetP|}{2}$ ; 
$ C = \sqrt[3]{\frac{\Delta_1 + \sqrt{\Delta_1^2 - 4\Delta_0^3}}{2}}$.
    If  both conditions $|DetP|=0$ and $TrP=\pm1$ are satisfied simultaneously then only the Schmidt decomposition has one term  and the state is unentangled, otherwise the state is entangled.	
    The entangled bipartite qutrit basis are
    $|\beta_0\rangle = \frac{|00\rangle+|11\rangle+|22\rangle}{\sqrt{3}} $ ,	
	$|\beta_1\rangle=\frac{|01\rangle+|10\rangle}{\sqrt{2}}$,
	$|\beta_2\rangle=\frac{|01\rangle-|10\rangle}{\sqrt{2}}$,
	$|\beta_3\rangle=\frac{|00\rangle-|11\rangle}{\sqrt{2}}$,
	$|\beta_4\rangle=\frac{|02\rangle+|20\rangle}{\sqrt{2}}$,
	$|\beta_5\rangle=\frac{|02\rangle-|20\rangle}{\sqrt{2}},$    $|\beta_6\rangle=\frac{|12\rangle+|21\rangle}{\sqrt{2}}$, $|\beta_7\rangle=\frac{|12\rangle-|21\rangle}{\sqrt{2}}$,
	$|\beta_8\rangle =\frac{|00\rangle+|11\rangle-2|22\rangle}{\sqrt{6}}$
$\langle \beta_i|\beta_j\rangle=\delta_{ij}$. $|\beta_0\rangle$ is the only maximally entangled state (reduced density matrix is $\frac{I}{3}$). All others are non-maximally entangled. The Gell-Mann matrices are:

	$	\lambda_1 = \begin{pmatrix}
			 0& 1 & 0 \\		  
			  1 & 0 & 0 \\
			   0 & 0 & 0
			\end{pmatrix},
		\lambda_2 = \begin{pmatrix}
			0 & -i & 0 \\ 		
			i & 0 & 0 \\
			 0 & 0 & 0 
			\end{pmatrix},
	\lambda_3 = \begin{pmatrix}
			1 & 0 & 0 \\
			 0 & -1 & 0 \\
			  0 & 0 & 0 
			\end{pmatrix},
		\lambda_4 = \begin{pmatrix}
			 0 & 0 & 1 \\ 
			  0 & 0 & 0 \\
			   1 & 0 & 0
			\end{pmatrix}$ \\
  $\lambda_5 = \begin{pmatrix}
			 0 & 0 & -i \\
			  0 & 0 & 0 \\
			   i & 0 & 0 
			\end{pmatrix},
		\lambda_6 = \begin{pmatrix}
			 0 & 0 & 0 \\  
			 0 & 0 & 1 \\
			  0 & 1 & 0 
			\end{pmatrix},
		\lambda_7 = \begin{pmatrix}
			0 & 0 & 0 \\
			 0 & 0 & -i \\
			  0 & i & 0
			\end{pmatrix},
		\lambda_8 = \frac{1}{\sqrt{3}} \begin{pmatrix}
			 1 & 0 & 0 \\
			  0 & 1 & 0 \\
			   0 & 0 & -2
			\end{pmatrix}$.
           
            If $|\chi\rangle_{AB}=|\beta_{\alpha}\rangle$ for $\alpha$=0,1,2..,8 ; then $a_{ij}^{(\alpha)}=\langle ij|\beta_{\alpha}\rangle$ and (i,j=0,1,2), then the P matrices can be represented in terms of $\lambda$'s for each of the entangled qutrit basis.\\       
        \begin{table}[H]
         \renewcommand{\arraystretch}{0.45} 
    \centering
    \begin{tabular}{|c|c|c|c|c|}
     \hline
        \textbf{$|\chi\rangle_{AB}=|\beta_{\alpha}\rangle$} & \textbf{Non-zero \(a_{ij}^{(\alpha)}\)} & \textbf{\(P_{\alpha}\)} & \textbf{\(Det P_{\alpha}\)} & \textbf{\(Tr P_{\alpha}\)} \\ 
        \hline 
        $|\beta_0\rangle$ & $a_{00}=a_{11}=a_{22}=1/\sqrt{3}$ &$\frac{1}{\sqrt{3}}I$  & $ 1/(3\sqrt{3})$ &  $\sqrt{3}$ \\
        \hline
         $|\beta_1\rangle$ & $a_{01}=a_{10}=\frac{1}{\sqrt{2}}$ & $\frac{1}{\sqrt{2}}\lambda_1$  & 0 &  $0$ \\
         \hline
          $|\beta_2\rangle$ & $a_{01}=-a_{10}=\frac{1}{\sqrt{2}}$ & $\frac{i}{\sqrt{2}}\lambda_2$  & 0 &  $0$\\ 
          \hline
           $|\beta_3\rangle$ & $a_{00}=-a_{11}=1/\sqrt{2}$ & $\frac{\lambda_3}{\sqrt{2}}$  & 0 &  $0$\\ 
           \hline
            $|\beta_4\rangle$ & $a_{02}=a_{20}=1/\sqrt{2}$ & $\frac{\lambda_4}{\sqrt{2}}$  & 0 &  $0$\\ 
           \hline
           $|\beta_5\rangle$ & $a_{02}=-a_{20}=1/\sqrt{2}$ & $\frac{i\lambda_5}{\sqrt{2}}$  & 0 &  $0$\\ 
           \hline
            $|\beta_6\rangle$ & $a_{12}=a_{21}=1/\sqrt{2}$ & $\frac{\lambda_6}{\sqrt{2}}$  & 0 &  $0$\\ 
           \hline
            $|\beta_7\rangle$ & $a_{12}=-a_{21}=1/\sqrt{2}$ & $\frac{i\lambda_7}{\sqrt{2}}$  & 0 &  $0$\\ 
           \hline
            $|\beta_8\rangle$ & $a_{00}=a_{11}=\frac{1}{\sqrt{6}};a_{22}=\frac{-2}{\sqrt{6}}$ & $\frac{\lambda_8}{\sqrt{2}}$  & $1/(3\sqrt{6})$ &  $0$\\ 
           \hline
    \end{tabular}
    \caption{The P matrices for entangled qutrit states}
  \label{qtritp}
\end{table}	
 \begin{table}[H]
 \centering
    \renewcommand{\arraystretch}{0.5} 
    \begin{tabular}{|c|c|c|c|c|c|}
    \hline
        \textbf{ijk} & \textbf{f\(_{ijk}\)} & \textbf{\(\theta(i,j)\)} & \textbf{ijk} & \textbf{f\(_{ijk}\)} & \textbf{\(\theta(i,j)\)} \\
        \hline
        123 & $f_{123}=+1$ & $\theta(1,2)=-1$ & 345 & $f_{345}=+1/2$ & $\theta(3,4)=+1$ \\
            &  & $\theta(2,3)=-1$ &  &  & $\theta(4,5)=-1$ \\
            &  & $\theta(3,1)=+1$ &  &  & $\theta(5,3)=-1$ \\
        \hline
        112 & $f_{147}=+1/2$ & $\theta(1,4)=+1$ & 367 & $f_{367}=-1/2$ & $\theta(3,6)=+1$ \\
            &  & $\theta(4,7)=-1$ &  &  & $\theta(6,7)=-1$ \\
            &  & $\theta(7,1)=-1$ &  &  & $\theta(7,3)=-1$ \\
        \hline
        113 & $f_{156}=-1/2$ & $\theta(1,5)=-1$ & 458 & $f_{458}=\sqrt{3}/2$ & $\theta(4,5)=-1$ \\
            &  & $\theta(5,6)=-1$ &  &  & $\theta(5,8)=-1$ \\
            &  & $\theta(6,1)=+1$ &  &  & $\theta(8,4)=+1$ \\
        \hline
        246 & $f_{246}=+1/2$ & $\theta(2,4)=-1$ & 678 & $f_{678}=\sqrt{3}/2$ & $\theta(6,7)=-1$ \\
            &  & $\theta(4,6)=+1$ &  &  & $\theta(7,8)=-1$ \\
            &  & $\theta(6,2)=-1$ &  &  & $\theta(8,6)=+1$ \\
        \hline
        257 & $f_{257}=+1/2$ & $\theta(2,5)=-1$ &  &  &  \\
            &  & $\theta(5,7)=-1$ &  &  &  \\
            &  & $\theta(7,2)=-1$ &  &  &  \\
        \hline
    \end{tabular}
    \caption{The non-zero structure constants and their associated $\theta(i,j)$ }
\end{table} 
The commutation relations between the $P$'s are:
\begin{equation}
     \label{non-redef-su3}
     [P_i,P_j]=\sqrt{2} \sum _k f'_{ijk}P_k ;\enskip and \enskip f'_{ijk}=\theta(i,j)f_{ijk}.
 \end{equation}  
It is easily noted that $P_2=-P_2^{\dagger} $ ; $ P_5=-P_5^{\dagger}$ and $ P_7=-P_7^{\dagger}$. Thus we can write $P_i=K_i^{\dagger}U_i$ where $i=2,5,7$ where $K_i^{\dagger}$'s are complex conjugation operators and $U_i$'s are unitary operators. We have seen similar results in section 2. This is a new result in the context of qutrits.

Redefining P matrices in table (\ref{qtritp}) as 
					$P_0'=\sqrt{3}P_0; P_1'=\sqrt{2}P_1 \enskip; \enskip P_2'=\sqrt{2}e^{(-i\pi/2)}P_2 	\enskip;\enskip P_3'=\sqrt{2}P_3 \enskip;\enskip 
				P_4'=\sqrt{2}P_4 \enskip;\enskip P_5'=\sqrt{2}e^{(-i\pi/2)}P_5\enskip;$ 
                 $P_6'=\sqrt{2}P_6\enskip;\enskip 
				P_7'=\sqrt{2}e^{(-i\pi/2)}P_7 \enskip;\enskip P_8'= \sqrt{2} P_8 $. We have 
			\begin{equation}
            \label{SU3}
				[P_{i}',P_{j}']=2i \sum_k f_{ijk}P_{k}'
			\end{equation}
			which is a SU(3) algebra where $f_{ijk}$'s are the usual structure constants.                 
 
\section{Entanglement entropy}
$Det A$ is related to the entanglement entropy $S$ which is a measure of the degree of entanglement (i.e. correlation) between the two subsystems. 
\begin{equation}	\label{entropy}
	S=-\mu_i ln(\mu_i)
\end{equation}
where Einstein's summation convention prevails. $\mu_i$ are the eigen values of the reduced density matrix of the composite state.

If two subsystems of the composite quantum system are not entangled (i.e. no correlation) then the entanglement entropy  $S=0$. For maximally entangled states  the entanglement entropy is maximum.

For a bipartite qubit state the four Bell states form maximally entangled basis states for which the entanglement entropy is $ln(2)$.  

\begin{figure}[h]
    \centering
    \includegraphics[width=0.75\linewidth]{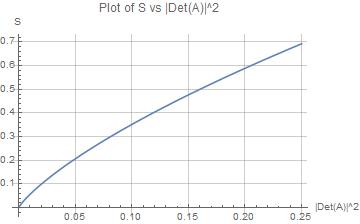}
    \caption{S vs $|Det A|^2$ for bipartite qubit system}
    \label{s vs d fig}
\end{figure}

The states for which $0<S<ln(2)$ are non-maximally entangled states.
The entanglement entropy for the general bipartite qubit state $|\psi\rangle$ is:
\begin{equation}
	\label{entropy-psi}
	\begin{split}
		S_{\psi}= 
        -(1/2)(1-\sqrt{(1-4|DetA|^2})ln((1/2)(1-\sqrt{(1-4|DetA|^2}))\\ \\
        -(1/2)(1+\sqrt{(1-4|DetA|^2)}ln((1/2)(1+\sqrt{(1-4|DetA|^2)})
	\end{split}
\end{equation}
The Fig (\ref{s vs d fig})shows that the entanglement entropy $S$ increases with  $|DetA|^2$ .

\begin{figure}
	    \includegraphics[width=1.20\linewidth]{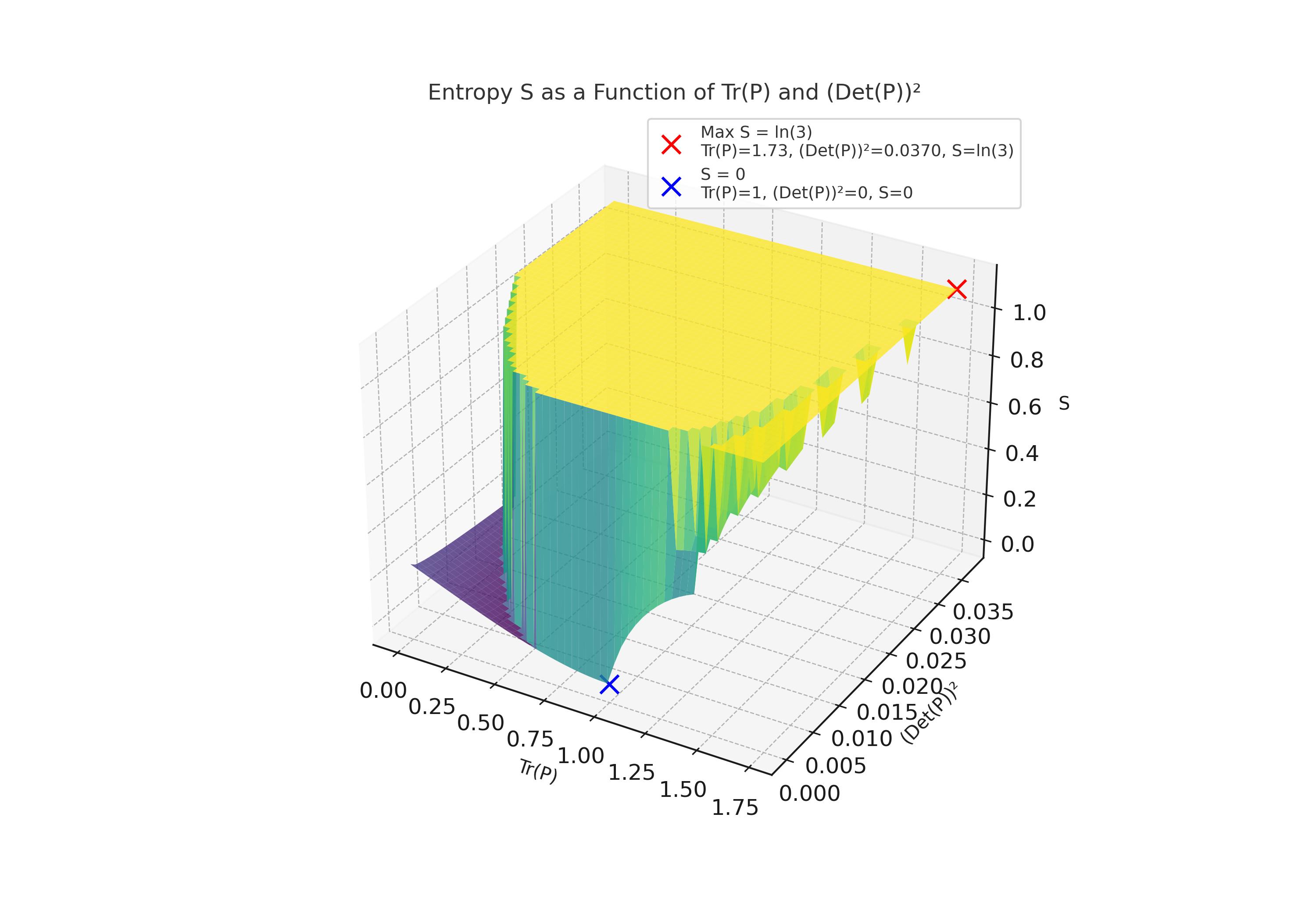}
	    \caption{Variation of Entanglement Entropy with TrP and $(DetP)^2$ for qutrits. }
	    \label{S qutrit}
	\end{figure}

For the maximally entangled bipartite qutrit system (for example $|\beta_0\rangle$) ,since the reduced density matrix corresponding to the state is $\frac{I}{3}$, the entanglement entropy $S=ln(3)$, and any state for which $0<S<ln(3)$ corresponds to non-maximally entangled bipartite qutrit states.The entanglement for the most general bipartite qutrit state $|\chi\rangle$ (equation \ref{chi}) is:
\begin{equation}
\label{S-chi}
    S_{\chi}=-\sum_{i=0}^2\mu_i ln(\mu_i)= -\sum_{i=1}^3(\frac{TrP}{3} + \frac{\omega^i C + \frac{\Delta_0}{\omega^i C}}{3})ln(\frac{TrP}{3} + \frac{\omega^i C + \frac{\Delta_0}{\omega^i C}}{3})
\end{equation}
where, $\mu_i=\frac{TrP}{3} + \frac{\omega^i C + \frac{\Delta_0}{\omega^i C}}{3}$, where $\Delta_0 , C$ and $\omega$ have already been defined in the previous section. The Figure \ref{S qutrit} shows how entanglement entropy for a bipartite qutrit system varies with Tr(P) and $(DetP)^2$.

        It can be easily seen from Figure \ref{S qutrit} that for the  unentangled state  S=0 (blue cross) corresponds to TrP=+1 and DetP=0  which is consistent with our results in the previous section. The red cross denotes one of the maximally entangled states with S=ln(3). Each point on the boundary of volume bounded by the green surface denotes a non-maximally entangled state. 

        Further, it is to be noted that each point on the yellow plane ({\it S=ln(3) }plane) corresponds to a maximally entangled bipartite qutrit state.
    \section{Conclusion:}
We have developed a host of analytic results related to entanglement of wavefunctions  for bipartite qubit and qutrit systems. We have also given a new protocol for teleportation using non-maximally entangled states with increased cryptographic security. We summarise as follows :
			
(a)For the bipartite qubit system there exists a matrix $A$, equation (\ref{A}), constructed from the amplitudes of the basis states. If $Det A = 0$, the state is unentangled. For $Det A \neq 0$ the state is entangled. Under a redefinition, the matrices $A$ satisfy an exact SU(2) algebra : sections 2 and 4 , equations: (\ref{A0}), (\ref{A1-3}) and (\ref{SU2}).

(b)Teleportation for a general qubit state is possible by using non-maximally entangled bipartite qubit states. This protocol has an additional parameter , {\it viz.},$Det A$, which enhances the cryptographic security of the teleportation : section 3. 

(c)Exhaustive physical interpretation  of the underlying algebras are given and plausible experimental scenarios are proposed for the SU(2) case in the context of two entangled electrons : sections 4 and 5.

(d)For the bipartite qutrit system one can again construct a matrix $P$, equation (\ref{P}), using the amplitudes $a_{ij}$ defined in equation (\ref{chi}). The entangled bipartite qutrit basis states are given  in Section 6. For the bipartite qutrit state  $Det P=0$ simultaneously with  $Tr P = \pm 1$ imply unentanglement. Any departure from these conditions imply entanglement. For the entangled qutrit system a SU(3) closed algebra is obtained using a redefinition of P matrices in terms of  the Gell-Mann matrices .This is demonstrated in equation (\ref{SU3}) using the relevant constructions as defined in Table (\ref{qtritp}). 
			
(e) Another interesting aspect follows from the discussions after table (\ref{theta(ij)qubit}) and equation (\ref{SU3}). The underlying SU(2) or SU(3) algebras were obtained only after 
converting certain antiunitary matrices to hermitian (and unitary) matrices by multiplying by  a factor of $\sqrt{2} exp[- i\pi /2]$, i.e. a unitary rotation by $\pi/2$.
Antiunitary matrices are related to time reversal. Recall that there is only one independent antiunitary symmetry whose physical meaning is  time reversal. Any other antiunitary transformation can be expressed in terms of time reversal, ({\it viz.}  a unitary matrix multiplied by time reversal). This is well known and was discussed in the literature$^{29}$. This fact has been built into our redefinition of the matrix $A_2$ as $A_ {2} '$ immediately after table (\ref{theta(ij)qubit}) for the qubit case. Similar redefinitions of $P_2 , P_5, P_7$ as $P_2 ', P_5 ', P_7 '$ respectively for the qutrit case have the same underlying physical meaning. Therefore, the unitary algebras underlying entangled states discussed here are somehow indicative of some time reversal scenarios. This aspect deserves further studies.

(f) Finally, we have shown how the entanglement entropy is related to $Det A$ for qubits and $Det P$ and $Tr P$ for qutrits.

\section{Acknowledgement}
One of the author(PDG) would like to thank Sister Nivedita University for providing research scholarship, under Student ID-2331207003001

\end{document}